\documentclass[11pt, tightenlines, onecolumn, amsmath, amssymb, notitlepage, a4paper, superscriptaddress]{revtex4-1}
\UseRawInputEncoding
\usepackage[strict]{revquantum}

\usepackage{graphicx}
\usepackage{algpseudocode} 
\usepackage{algorithm}
\usepackage{subfig}
\usepackage{braket}
\usepackage[margin=1in]{geometry}
\usepackage{bbm}

\DeclareMathOperator{\tr}{tr}

\algnewcommand\algin{\textbf{in}}
\DeclareMathOperator*{\argmax}{arg\,max}

\begin{document}

\title{Noise Detection with Spectator Qubits and Quantum Feature Engineering}

\author{Akram Youssry}
\email[Corresponding Author]{}
\affiliation{Quantum Photonics Laboratory and Centre for Quantum Computation and Communication Technology, RMIT University, Melbourne, VIC 3000, Australia}
\affiliation{Centre for Quantum Software and Information, University of Technology Sydney, Ultimo NSW 2007, Australia}
\author{Gerardo A. Paz-Silva}
\affiliation{Centre for Quantum Dynamics and  Centre for Quantum Computation and Communication Technology, Griffith University, Brisbane, Queensland 4111, Australia} 

\author{Christopher Ferrie}
\affiliation{Centre for Quantum Software and Information, University of Technology Sydney, Ultimo NSW 2007, Australia}

\date{\today}

\begin{abstract}
Designing optimal control pulses that drive a noisy qubit to a target state is a challenging and crucial task for quantum engineering. In a situation where the properties of the quantum noise affecting the system are dynamic, a periodic characterization procedure is essential to ensure the models are updated. As a result, the operation of the qubit is disrupted frequently. In this paper, we propose a protocol that addresses this challenge by making use of a spectator qubit to monitor the noise in real-time. We develop a machine-learning-based quantum feature engineering approach for designing the protocol. The complexity of the protocol is front-loaded in a characterization phase, which allow real-time execution during the quantum computations. We present the results of numerical simulations that showcase the favorable performance of the protocol.  

\end{abstract}


\maketitle
%\tableofcontents

%%%%%%%%%%%%%%%%%%%%%%%%%%%%%%%%%%%%%%%%%%%%%%%%%%%%%%%%%
\section{Introduction}
As quantum technology progresses more towards the Noisy Intermediate-Scale Quantum (NISQ) devices era \cite{preskill2018quantum}, the design and operation tasks become more challenging. One such task is quantum control, where the objective is to perform a desired operation with the best possible fidelity (or at least above a desired standard) given available controls. Many techniques have been developed in this space. On the one hand, one has model-uncertainty robust techniques, such as dynamical decoupling or dynamically-corrected gates \cite{OC1, OC2, DS1, DS2, DS3, DS4, DS5}, which can perform a gate with relatively minimal assumptions on the noise, e.g., that it is mainly low-frequency. On the other, one has optimal control (OC) techniques \cite{PhysRevA.99.052327, machnes2015gradient, Ciaramella_2015, caneva2011chopped, de_Fouquieres_2011, khaneja2005optimal}, which use numerical optimization to find the control sequence that achieves a desired operation given assumed knowledge of the noise. A clear trade-off exists between assumed knowledge of the bath and effectiveness of the control scheme, i.e., the less robust to model uncertainty is the technique the better its performance is and vice versa. 

A way to get the best of both worlds is to approach the problem in two stages. The first is a characterization stage, which aims to produce models of the noise based on experimental measurements. Quantum Noise Spectroscopy (QNS) is one example where the Power Spectral Density (PSD) of the noise -- or the higher versions of these depending of the generality of the protocol -- is estimated by measuring the response of the quantum system of interest to various control routines while in the presence of the noise to be characterized~\cite{qns1, qns3, qns4, ferrie2018bayesian, PhysRevLett.107.230501, szankowski_environmental_2017, krzywda_dynamical-decoupling-based_2019, frey2019, Frey2017, NonGaussianExp, NonGaussianPRL, Ramon, multiaxis, multiqubit, LukazOpt, PhysRevA.98.032315, added2, added4, added5, added6, added7}. In the second stage OC techniques are applied and a cost function {\it based on the acquired information in the first stage} is optimized with respect to the control. 

Recognizing the importance of characterizing open quantum systems, quantum noise spectroscopy techniques have become now prevalent. A common characteristic of these protocols is that they require expressions for the qubit dynamics, e.g., of expectation values of observables, in terms of quantities describing the bath, e.g., bath correlations. Since rarely one can have a closed from expression for these quantities and one requires some form of perturbation theory, it follows that QNS protocols are necessarily perturbative. The truncation in perturbation theory is usually justified by a weak coupling or short time assumption, and thus regimes in which the truncation conditions are not satisfied, cannot be addressed. Additionally, many of theses techniques assume ideal control pulses (for example instantaneous or unlimited bandwidth). In practice, the non-idealities will affect the performance of the device if not accounted for during the design process. Moreover, until very recently, these protocols were agnostic to the implications on the learnable information the bath given fixed control constraints, i.e., given fixed control capabilities only so much about the noise can be learned.
 
A Machine Learning (ML) approach has been proposed recently to address these drawbacks in \cite{BQNS,youssry2022multi, Youssry_2020, youssry2022experimental}, the so-called Graybox approach to control. Namely, it is non-perturbative which naturally takes into account control constraints. The idea is to construct graybox  structures that consists of standard blackbox layers (such as Neural Networks (NN)), as well as custom whitebox layers that encode quantum operations (such as quantum evolution). The advantage of this approach is that it allows having assumptions-free models because of the use of the blackbox layers. At the same time, it allows estimating significant physical quantities (such as Hamiltonians), satisfying their mathematical constraints (such as Hermiticity) due to the use of whiteboxes. The models are trained from experimental measurements and can then be used to design the control. 

\begin{figure}[]
    \centering
    \includegraphics[scale=1]{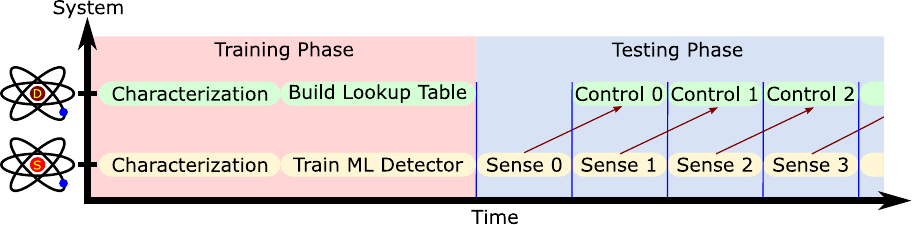}
    \caption{Timing diagram of the proposed protocol. The upper qubit labeled `D' is the data qubit, and the lower qubit labeled `S' is the spectator qubit. At the beginning there is a training phase that is executed once which includes characterization of the qubits and training the machine learning models. This is followed by the testing phase in which a periodic cycle of measurements of the spectator qubit inform the optimal control of the data qubit.}
    \label{fig:timing}
\end{figure}

These methods have been successfully applied to systems over fixed evolution times, showing that one can indeed learn about the systems enough to control it over such time. The caveat of these methods is their scaling with evolution time. The longer the evolution time the more costly the learning and the optimization becomes. An exception to this crippling scenario is when the noise is stationary, i.e., its statistics is invariant under time translations. That is, every time one runs an experiment the statistics of the noise are assumed to be the same or, equivalently, the statistics of the noise are time translation invariant. In such scenario, the `time-local' characterization can be used at any point in time because of the aforementioned invariance. 

Unfortunately, while this may hold in relatively short timescales, it is not true in general. Indeed, noise is generally non-stationary, even if it can be considered approximately stationary at short time scales. This is especially true when noise is generated by a quantum environment. Consider, for example, the effect of the noise generated by fluctuators in a solid-state system. When the position of the charge in each fluctuator changes, the noise affecting the qubit of interest changes, i.e., the noise can be considered non-stationary in a long-time scale but stationary in a short timescale. The usual routine is to periodically recalibrate the system to accommodate for this, but this is clearly not an ideal solution, as this implies stopping the computation and thus limiting the total usable coherent time of the qubit. 

One solution to this problem is through the use of spectator qubits \cite{PhysRevA.102.042611, Majumder_2020, Song_2023, Tonekaboni_2023}. Rather than considering the non-stationary noise over a long time scale, we ``coarse-grain'' the problem by considering that along its evolution the qubit can be subject to a finite set noise of profiles, e.g., there is a finite number of noise fluctuators in the solid-state example above. The idea is then to use another qubit, the ``spectator'', as a sensor to the active noise configuration so that there is no need to stop an experiment, e.g., a  quantum computation, on the main ``data'' qubit to recalibrate. The spectator can be a ``lower-quality'' qubit in the sense that it is better if it is more sensitive to the noise, but one requires that one has at least the same control capabilities as the ``data'' qubit. The latter will be an important consideration, as we will see later. Figure \ref{fig:timing} shows a typical timing diagram for such a protocol. Following the reasoning above, in this paper we then address the problem of designing a sensing protocol utilizing a spectator qubit for noise detection \cite{martina2021machine, scholten2019classifying}. We will utilize a ``Quantum Feature Engineering'' (QFE) approach, extending the proposal in \cite{BQNS} and drawing analogy from the paradigm of ``feature extraction'' in classical machine learning.

The rest of this paper is organized as follows. Section \ref{sec:problem} describes the problem setting and the underlying assumptions. Next, Section \ref{sec:methods} introduces details of the protocol as well as the design of the ML models it utilizes. After that, we give details on the numerical simulations we implemented to verify the proposed ideas in section \ref{sec:sim}. Finally, we conclude paper and show the potential extensions of the presented work in Section \ref{sec:conc}. 

\section{Problem Setting} \label{sec:problem}
The basic setup of our problem is as follows. The data qubit is subject to a Hamiltonian of the form
\begin{align}
    H_D(t) = H^{(\text{ctrl})}_D(t) + H_D^{(1)}(t). 
\end{align}
The first term is known and deterministic, and represents the drifting and control Hamiltonians given by  
\begin{align}
    H^{(\text{ctrl})}_D(t) = \frac{1}{2}\Omega_D \sigma_z + \frac{1}{2} \sum_{k\in \{x,y,z\}} f_D^{(k)}(t) \sigma_k,
\end{align}
where $\Omega_D$ is the energy gap of the qubit, $\sigma_k$ are the Pauli matrices, and $f_D^{(k)}(t)$ is the control pulse sequence driving the qubit along the $k^{\text{th}}$ direction. On the other hand, the second term $H_D^{(1)}$ is an unknown stochastic Hamiltonian that encodes the noise affecting the qubit due to interaction with the environment. We can express it generally in the form
\begin{align}
    H^{(1)}_D(t) = \sum_{k\in \{x,y,z\}} \beta_D^{(k)}(t) \sigma_k,
\end{align}
where $\beta_D^{(k)}(t)$ is a realization of the noise random process $\mathbf{B}_D(t) $ along the $k^{\text{th}}$ direction. 
In other words in a given time interval, the realization of $\beta_D^{(k)}(t)$ is taken from a noise process described by one of $N$ possible set of correlation functions $\{\langle \beta_D^{(k)}(t)\rangle_c^{(i)} , \langle \beta_D^{(k)}(t)\beta_D^{(k)}(t')\rangle_c^{(i)}, \cdots \}_{i=1,\cdots N },$ where $\langle \cdot \rangle_c$ denotes average over noise realizations. Similarly, the spectator qubit is subject to a Hamiltonian of the form,  
\begin{align}
    H_S(t) = H^{(\text{ctrl})}_S(t) + H_S^{(1)}(t),  
\end{align}
where 
\begin{align}
    H^{(\text{ctrl})}_S(t) = \frac{1}{2}\Omega_S \sigma_z + \frac{1}{2} \sum_{k\in \{x,y,z\}} f_S^{(k)}(t) \sigma_k.
\end{align}
The spectator qubit energy gap $\Omega_S$ and the control pulses $f_S^{(k)}(t)$ need not be the same as the data qubit, and thus the noise Hamiltonian of the spectator can also be written in the form
\begin{align}
    H^{(1)}_S(t) = \sum_{k\in \{x,y,z\}} \beta_S^{(k)}(t) \sigma_k, \label{equ:H1}
\end{align}
where $\beta_S^{(k)}(t)$ is a realization of a random process $\mathbf{B}_S(t) $ along the $k^{\text{th}}$ direction. 

Information about a quantum state is extracted via measurements. Concretely, via the expectation value $\braket{O(T)}$ of some observable $O$ on subsystem $A \in \{S,D\}$ at time $t=T$. Assuming the qubit starts in the state $\rho_A(0)$, using a modified interaction picture (see \cite{BQNS} for a detailed derivation), we can write
\begin{align}
    \braket{O(T)} = \tr{\left(V_O(T) U_0(T) \rho_A(0) U_0^{\dagger}(T) O \right)}, \label{equ:O}
\end{align}
where %
\begin{align}
     U_0(T) = \mathcal{T}_{+} e^{-i \int_0^T H^{(\text{ctrl})}_A(t) dt }, \label{equ:U0}
\end{align}
represents the evolution of the system in the absence of noise. In contrast, the operator $V_O(T)$ encodes all information about the noise and how it interacts with the control, and is given by
\begin{align}
    V_O(T) = O^{-1}\braket{\tilde{U}_I(T)^{\dagger}O\tilde{U}_I(T)}_c \label{equ:Vo},
\end{align} with 
\begin{align}
    \tilde{U}_I(T) = U_0(T) U_I(T)U_0^{\dagger}(T), \label{equ:UI_tilde} 
\end{align}
and 
\begin{align}
    U_I(T) = \mathcal{T}_{+} e^{-i \int_0^T U_0^{\dagger}(t) H^{(1)}_A(t) U_0(t) dt}.
\end{align}

The $V_O$ formalism allows us to separate the effect of noise on the expectation value of some observable, i.e., in the absence of noise $V_O$ is the identity. This is crucial for the success of the graybox approach since, in principle, it is a non-linear functional of the control and this machine learning is ideally suited to estimate it, by measuring the response of measured expectation values to various controls.

%In principle, if enough information about the noise statistics is available, say obtained via QNS~\cite{}, then optimal control techniques~\cite{} can be used to implement a desired guide with high fidelity. However, if noise statistics change, then the optimal solution may no longer apply.  

\subsection*{Noise assumptions and control-dependent open quantum system characterization} 

In this paper, we focus on the setting in which there is a common physical mechanism inducing stochastic noise on both the data and spectator qubits, described by some set of parameters $\Theta$. We are interested in the scenario where the configuration of $\Theta$ changes in principle discretely over a certain timescale, leading to changes the statistics of the data and spectator noise. We will assume that there are $N'$ possible non-equivalent configurations of $\Theta,$ and, consequently, the same number of noise profiles affecting each qubit, i.e.,
  \begin{align}
    \mathbf{B}_A(t) \in \Big\{  \mathbf{A}_k \Big\}_{k=1}^{N'}. \label{equ:Bset}
\end{align}
for $A \in \{S,D\}.$

A key element in our protocol will be the graybox characterization of open quantum systems~\cite{BQNS}. The approach generates a control capability ($\mathcal{C}$)- dependent map $G$ between the noise affecting the system and a neural network representation of it, i.e.,
$$
\mathcal{G}(\mathcal{C}): \mathbf{A}_k \rightarrow \mathbf{G}_{A_k; \mathcal{C}} 
$$
for $A \in {S,D}.$ The training data necessary to achieve this consists of expectation values of a complete set of operators and initial states under a sufficiently large set of control configurations within $\mathcal{C}.$ Now, given $\mathbf{G}_{A_k; \mathcal{C}}$ one can then use optimal control to achieve any desired task on $A$ via an appropriate cost function, e.g., an optimal fidelity gate as we did in~\cite{BQNS}, provided the same constraints are use in the the search for the optimal solution. We highlight that the $\mathcal{C}$-dependence of the map is at the core of graybox's success. Not only is it what leads to its efficiency (see also ~\cite{Behnam20} for a formal treatment of the argument) but, particularly relevant to the present work, it implies that two noise models are effectively  different as long as the available control can discriminate them, i.e., only noise models distinguishable under the available control matter. In other words, even if the noise is generated by $N'$ possible configurations one is only interested in $N$ sets of configurations that are distinguishable via $\mathcal{C}$. 

We will crucially assume that the graybox protocol is run simultaneously on both the data and spectator qubit over a long time, and that during this period the noise will cycle over the $N'$ profiles. Further we will assume that the model stays sufficiently long in a given profile to allow for a satisfactory graybox characterization of each profile. Because the control capabilities on both spectator and data qubit are assumed to be in principle different, each qubit can distinguish between different types of profiles, say $M$ and $N$ respectively. If $M\geq N$, it is possible to build a {\it surjective} map $\mathcal{F}$, between the noise profiles affecting the spectator qubit and those affecting the data qubit, 
\begin{align}
    \mathcal{F}: \Big\{  \mathbf{G}_{\mathbf{S}_k;\mathcal{C}} \Big\}_{k=1}^{M} \to \Big\{   \mathbf{G}_{\mathbf{D}_k;\mathcal{C}} \Big\}_{k=1}^{N}.
\end{align}
Notice that we are not assuming that data and spectator feel the same noise. Rather, that the noise they feel is somehow ``correlated". That is, if we assume that the noise is generated by a physical process described by a set of parameters, say $\Theta$, then $\beta_D(t)$ and $\beta_S(t)$ are a (potentially different) function of a possible $N'$ configurations of such parameters. For example, $\beta_D (t) = (\beta[\Theta](t) )^2$ while $\beta_S(t) = \beta[\Theta](t) $ for some process $\beta[\Theta](t).$ The assumption then is that the noise profile the data qubit is undergoing at a given time can be inferred by appropriately interrogating the spectator. For $\mathcal{F}$ to be surjective it is enough to demand that the spectator is a copy of the data qubit (in which case is injective), but generally one requires that the spectator (under its associated control) is at least as sensitive to the noise as the data qubit. This assumption can be motivated by studying for instance a spin-boson model with multi-qubits sharing the same quantum bath (see \cite{Breuer_2007} for detailed analysis). If the qubits are located in close proximity of each other, i.e. the minimum distance between the qubits is less than the correlation length of the bath, the coupling constants each qubit to the bath modes will be identical. Effectively, they sense the same noise. Another possible scenario is when the data/spectator qubits are exposed to a constant (realistically stochastic varying around a mean value) magnetic field which is not necessarily narrowly localized and each qubit couples to it with a different strength. This last scenario could materialize in an NV or quantum dot experiment. For the purposes of this paper, we will assume the $\mathcal{F}$ map is given, and we leave its detailed construction for a separate work.
 
    In summary, our assumptions can be distilled as:
\begin{enumerate}
    \item The number of control-distinguishable profiles ($M,N$) is fixed.
    \item The statistical properties associated to each $\Theta$ configuration are fixed (can be unknown).
    \item The measurements and control are fast enough such that the noise profile does not change during measurements or during execution of a quantum gate.  
     \item It is in practice possible to characterize each noise profile at the beginning of the protocol.
    \end{enumerate}

The first two assumptions are necessary in any signal detection problem, that is the existence of fixed well-defined classes to which the signal can belong. The third assumption is fair. If the noise switches between two profiles faster than the measurement or control time, then they should not be treated as separate profiles, but rather one profile with non-stationary statistics. Note that any kind of statistics are allowed for the profiles $\mathbf{A}_k$. The last assumption implies the ability to run a pre-characterization on all noise profiles. That is, the ability to gather statistics over a sufficiently long time so that the possible set of control-distinguishable noise profiles can be identified and characterized via graybox. Finally, we note here that if the map $\mathcal{F}$ changes over time, then it is an indication that a significant change in the qubits and/or the environment has taken place and a pre-characterization has to be implemented again. We note that with the current status of quantum technology, any device needs calibration after operating for an extended period of time. Therefore, constructing the map $\mathcal{F}$ should be a part of that procedure.

So, under the aforementioned settings and assumptions, we propose the following research question:\\

\textit{Can we design a protocol capable of applying an optimal gate (on average) on the data qubit at any time?} \\

The answer to this question is affirmative. It revolves around the idea of using the spectator qubit as a probe to the noise, while never measuring the data qubit during a computation. We will develop a protocol that utilizes ML techniques to model the qubits, design the optimal control pulses, and design a noise discriminator. In the next section, we will explain the protocol in detail.

\section{Methods}\label{sec:methods}
The protocol developed here is based on drawing an analogy between the noise detection problem (i.e. interrogating the spectator qubit to detect the current noise model) and the classification problem in ML (i.e. assigning labels to signals). While the standard approach in ML treats signals in an abstract sense, the noise detection problem needs to be treated differently to account for the underlying physics and constraints (e.g. what quantities can be accessed experimentally). Therefore, we customize a new ML approach that considers the physics of the problem and refer to as ``quantum feature engineering'' approach. Since this idea draws heavily from the classical ML problem, we start in Section \ref{sec:QFE} by giving a quick overview on some important notions used in ML, and how we shall define similar notions for the ``physics-aware'' approach. Next, we summarize the main stages of the protocol in Section \ref{sec:protocol}, followed by a detailed description of each stage in Sections \ref{sec:stage1} to \ref{sec:stage4}.

\subsection{Quantum feature engineering}\label{sec:QFE}
\subsubsection{Overview on classification in ML}
In classical ML literature, the term ``features'' refer to vectors extracted from a signal, that can be used for a variety of applications including classification (assigning labels to signals). Features can be ``raw'' such as the color of a pixel in an image or the amplitude of an audio signal or can be more abstract like the amplitude of a frequency component of a signal. There are three basic steps to prepare the features for a classifier:
\begin{enumerate}
    \item Feature generation: computing the feature vector from a given signal. 
    \item Feature selection: choosing a subset that best distinguishes the objects we are classifying, based on some ranking criterion. 
    \item Feature extraction:  applying transformations to enhance the distinguishability between the classes and also to reduce the dimensionality of the feature vector.
\end{enumerate}
Once these steps are done, an ML blackbox structure is constructed to perform the classification process. The structure is trained by minimizing a loss function (such as the mean-square error (MSE)) with respect to the model parameters over a training set. The loss function captures the error between the predicted label by the model, and the true label, for a given set of training examples. A training set thus consists of representative examples from each class, where an example is a pair of feature vector and the corresponding ground truth label. After the classifier is trained, it can then be used to classify new examples that are not part of the training set. This is referred to as the testing stage of the classifier. The training is usually a computationally expensive process, whereas the testing is very efficient. Traditionally, the process of feature generation, selection, and extraction are done manually using heuristics (see \cite{10.5555/1200914} for a standard text on the subject and \cite{Youssry_2016} for an application in image processing). On the other hand, in the modern paradigm of deep learning, these steps become an integrated part of the ML structure, and the inputs directly become the raw signal. The structure includes trainable layers that generates and extracts the features automatically as a part of the training process. The algorithm learns the optimal features that need to be generated and how to transform them such that the loss function is minimized. An example of this structure is the Convolutional Neural Network (CNN). 

\subsubsection{Analogy between classification and noise detection}
The noise detection problem can be formulated using the language of features and classification. Since the underlying physical model is based on quantum mechanics, we will refer to this approach as ``quantum feature engineering'' (see Table \ref{tab:analogy} for the analogy with standard ML). We need to train a classifier using features that encode information about the noise. The first step is then to identify an appropriate object from which such information can be experimentally extracted. 

Let us start by recognizing that information can be accessed via measurements and in particular the expectation value of some observables. According to Equation \ref{equ:O}, these depend on the noise, control (including evolution time), initial state, and observable. Importantly, however, we have already established that the effect of these various factors can be separated and that the $V_O(T)$ operator condenses the effect of noise and its interaction with control, relative to a measurement of the observable $O$ . Any observable that one measures can be written as a linear combination of elements of an appropriate basis, and thus by choosing an informationally complete set of observables and initial states it is possible to access all $\{ V_O (T)\},$ which justifies our choice of measurements as features in the first place. For a qubit, this would be six eigenstates of the Pauli operators as initial states, and all the Pauli operators as observable, which gives a total of 18 measurements. However, in certain cases we can reduce this set for efficiency purposes. For example, if we know that the noise is pure dephasing along Z-axis in the toggling frame associated to a fixed $\mathcal{C}$, then it is sufficient to measure the Pauli X operator, with the three positive eigenstates of the Pauli operators as initial states. This gives a total of 3 measurements. This step is analogous to feature selection in ML. 

Finally, one has to consider the effect of control, as it is what allows us to resolve different noise setups. Indeed, given a fixed control sequence, there is an infinite number of noise profiles capable of generating the same expectation value data. For example, in the case of ideal instantaneous pulses, and Gaussian stationary zero-mean dephasing noise, the measurement $\braket{X(T)} = e^{-\int S(\omega)|F(\omega)|^2 d\omega} \braket{X(0)}$, where $S(\omega)$ is the PSD of the noise, and $|F(\omega)|^2$ is the filter function. Clearly, for any filter there exists an infinite number of $S(\omega)$ leading to the same value of the exponent and thus to $\braket{X(T)}$. What is more, given a set of control capabilities $\mathcal{C}$ capable of generating multiple (potentially infinite) sequences, it can also be the case that multiple spectra are experimentally indistinguishable. For example, if control generates mostly generates filters supported at low frequency, then spectra which differ only in their high frequency components are indistinguishable. Therefore, a fixed $\mathcal{C}$ generates equivalence classes in the noise, in parallel to the point we made earlier on how the graybox approach generated a $\mathcal{C}$-dependent characterization of the noise. 

It follows then that the noise profiles can be related to the corresponding $V_O$ operators for each subsystem. Concretely, to each noise profile one can associate a highly non-linear function  $\tilde{V}^{(i)}$ which takes as inputs the control applied and observable measured and generates the corresponding $V_O$ operator encoding the measurable effect of the noise relative to $O$. Now, given fixed control capabilities the $\tilde{V}^{(i)}$ can be indistinguishable from each other, in the sense that any control in $\mathcal{C}$ yields the same expectation values for two distinct noise setups. Thus we have the set 
\begin{align}
V_{A: \mathcal{C}} \in \left\{ V^{(k)} \right\}_{k=1}^{M},
\end{align}
of distinguishable noise profiles 
where each $V^{(k)}$ corresponds to a set of profiles indistinguishable via $\mathcal{C}$ on qubit $A$. This highlights the importance of  graybox in our protocol: it in essence a ML-based emulator for the $V^{(m)}$ which can be used off-line to generate optimal control strategies.

\subsection{The proposed protocol}\label{sec:protocol}

Our protocol will be divided in two phases: training and testing. In each phase, the role of the spectator and data qubit will be different. 

{\it Training phase.-} The training pipeline starts with the characterization information as a raw signal, from which the $V_O$ operators for both the data and spectator are generated. An optional step would be feature selection where we select the combination of observables and initial states that will maximize the information obtained from the measurements (if we have some prior information about the noise). This characterization takes the form of a graybox representation of the $V_O$ for both the data and spectator qubit. With this, one can build the surjective function $\mathcal{F}$ identifying characterizations in the spectator with the ones in the data qubit, as discussed earlier. Moving forward, the characterization plays a different role in the data and spectator qubit.

On the data qubit, it allows one to build a lookup table $\mathcal{L}$ indexed by the noise profile and a desired quantum gate. In other words, a classical optimizer is run that outputs the control pulses $\mathbf{f}_D(t)$ which, when applied to the noise profile indexed by $n_D$, implement a desired gate $G$ with the best possible fidelity allowed by $\mathcal{C}$, i.e.,  
\begin{align}
    \mathbf{f}_D(t) = \mathcal{L}(n_D, G).
\end{align}
Notice that in principle only a universal set of gates is sufficient. However, a larger table will allow for better performance overall in a computation as it allows one to compile circuits more efficiently, i.e., the noisy implementation of $G = G_1 G_2$, say $\tilde{G}$, is generally better than that implementing $\tilde{G_1} \tilde{G_2}$. The details about how to construct such a table is out of scope of this paper, however the general method in \cite{BQNS} could be used by performing it for every possible noise profile. 

On the spectator qubit, the graybox characterization allows the design of a control sequence $\hat{\mathbf{f}}(t)$ (although could be multiple ones) and a set of observables which ensure maximum separation between classes (i.e. maximize the distinguishability between $V^{(k)}$ for the given control constraints), dubbed $C_D$. This is the feature extraction step. With this, a classifier is trained using the measurement data and the correct label of the noise profile. The outcomes of this pipeline will be the lookup datel $\mathcal{L}$ are (i) the $C_D$ and (ii) the trained classifier. 

The training pipeline will have a long execution time, however it is done once at the beginning of the protocol.

{\it Testing phase.-} The testing pipeline during which the device operates is more efficient. It starts by measuring the observables corresponding to $C_D$. The measurements are then passed to the trained classifier to predict the current noise profile affecting the system. Once a noise profile has been identified, the correct optimal gate implementation is chosen from the prebuilt lookup table and the desired gate is implemented with the best possible fidelity allowed by $\mathcal{C}$. This process will be repeated periodically over the time. The data qubit is never interrupted while executing the gates, which is the main objective of this paper.

%%%%%%%%%%%%%%%%%%%%%%%%%%%%
\begin{figure}[]
    \centering
    \includegraphics[scale=0.75]{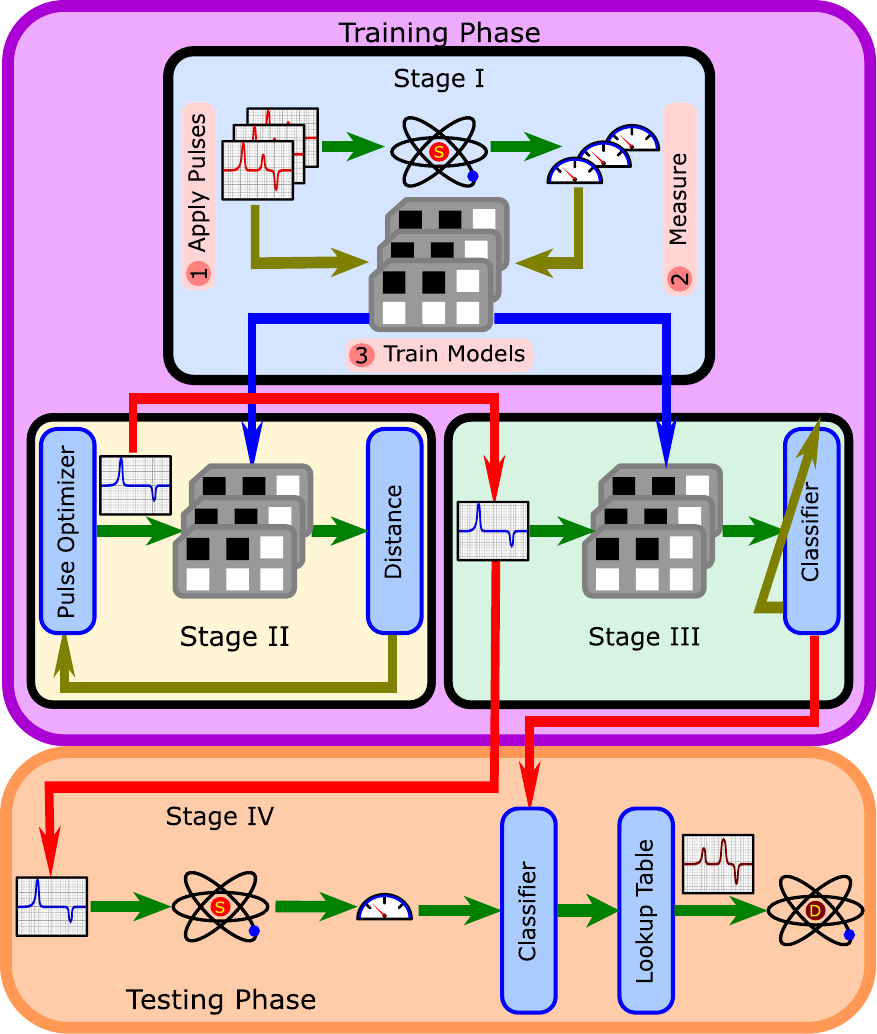}
    \caption{The proposed protocol for controlling a data qubit using measurements from a spectator qubit and quantum feature engineering. The first stage is for characterization and training ML models for estimating the $V_O$ operators. The second stage is using quantum control and the trained ML models to optimize the $V_O$ operator to maximize the distinguishablity of the classes. The third stage is training a classical classifier for detecting the noise profile. These three stages form the training phase of the protocol and are executed once at the beginning. The testing phase of the protocol is the actual sense-control periodic cycles. The measurements of the spectator corresponding to the optimal control sequence are used fed into the trained classifier. The predicted label is then used to lookup the optimal sequence that controls the data qubit given the current noise profile.}
    \label{fig:protcol}
\end{figure}

\subsection{The protocol}
We now flesh out the key steps in each phase. Note that the remainder of this paper, including the numerical simulations, will focus on the spectator qubit, and how the different steps are implemented. The protocol components related to the data qubit are standard tasks (characterization and control) and built in detail in \cite{BQNS}, and so we will not focus on these aspects in this paper.

 We will group the different steps of the protocol into four stages. The first stage (Section \ref{sec:stage1}) includes the characterization and estimation of the $V_O$ operators. We are going to use an ML approach as in \cite{BQNS}, but with a different design. The second stage presented in Section \ref{sec:stage2} will include the optimal control pulse design. In Section \ref{sec:stage3}, we explain the third stage which is training the classifier. Those three stages represent the training phase of the protocol. The last stage represents the testing phase of the protocol and is presented in Section \ref{sec:stage4}. The proposed protocol is summarized in Figure \ref{fig:protcol}.
\begin{table}[]
    \centering
    \begin{tabular}{|c|c|}
         \hline
         \textbf{ML} & \textbf{QFE}\\
         \hline
         Classes & Noise profiles \\ 
         \hline
         Raw signal & Characterization data \\
         \hline
         Feature generation & Estimating $V_O$ \\
         \hline
         Feature selection & Selecting best observables and initial states \\
         \hline
         Feature extraction & Designing the optimal control pulses \\
         \hline
         Features & Predicted optimal measurements\\
         \hline
    \end{tabular}
    \caption{The analogy between the proposed quantum feature engineering approach for addressing the noise detection problem, and classification}
    \label{tab:analogy}
\end{table}

\subsubsection{Stage I: Feature Generation}\label{sec:stage1}
\begin{figure}
    \centering
    \includegraphics{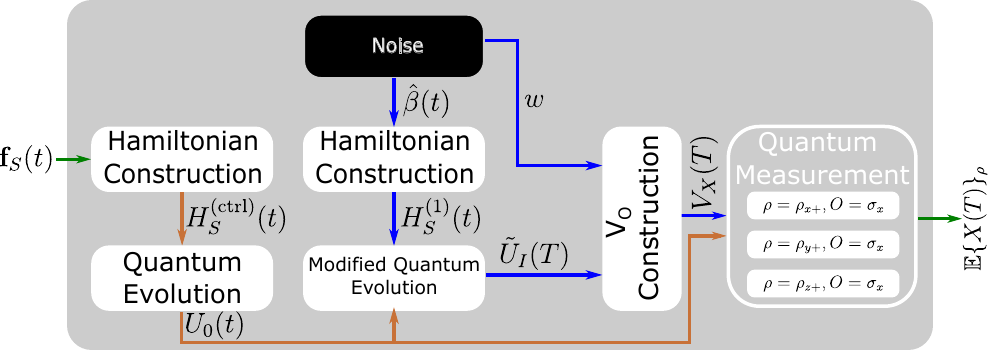}
    \caption{The proposed graybox structure for modeling a qubit. The input to the model is the control pulse sequence, and the output is the observables. The structure consists of two paths. The first path is the control path which starts with the time domain representation of the control pulse sequence as an input, followed by the construction of the control Hamiltonian and unitary. The second is the noise path that starts with a customized blackbox that generate noise realizations and weights, followed by the construction of the modified interaction picture unitary and finally the $V_O$ operator. The two paths merge at the output layer which calculates quantum observable parameterized by the initial state of the qubit.}
    \label{fig:mlmodel}
\end{figure}

The focus in this paper is on the spectator qubit. We will construct a graybox ML structure \cite{Youssry_2020, BQNS} to model the qubit. It consists of whitebox layers that performs quantum calculations and is able to generate features such as the $V_O$ operators. Additionally, it has blackbox layers that can be trained to generate information about the noise for example. The combination of blackboxes and whiteboxes result in an overall graybox. The proposed structure is shown in Figure \ref{fig:mlmodel}. The details are given next.

There are two main paths in the proposed ML structure. The first path is the control path. It starts with the model input which is the control pulse sequence $\mathbf{f}_S(t)$ represented in time domain. The control pulses then passes through a whitebox layer that constructs the Hamiltonian $H^{(\text{ctrl})}_S(t)$. After that, there is a whitebox that computes the control unitary $U_0(t)$ by approximating the time-ordered evolution in Equation \ref{equ:U0} as 
\begin{align}
    U_0(t) = e^{-i  H^{(\text{ctrl})}_S(M\Delta t) \Delta t } \cdots e^{-i  H^{(\text{ctrl})}_S(2\Delta t) \Delta t } e^{-i  H^{(\text{ctrl})}_S(\Delta t) \Delta t }  e^{-i  H^{(\text{ctrl})}_S(0) \Delta t }, \label{equ:timeorder}
\end{align}
where $\Delta t = t/M$, and $M$ is the number of discrete time steps.  

The second path is the noise path. It starts with a custom blackbox that has two outputs. The first is a set of normalized weights $\{w_k\}_{k=1}^{K}$, such that $0\le w_k \le 1$, and $\sum_k w_k = 1$. That is, they form a probability distribution. These weights are trainable, so during the training process, the loss function is optimized with respect to those weights. In order to ensure that the conditions hold, we can simply start with a general unconstrained set of weights $\{\tilde{w}_k\}_{k=1}^{K}$, and then pass them to a standard softmax activation layer that implements the transformation
\begin{align}
    w_k = \frac{e^{\tilde{w}_k}}{\sum_{k=1}^{K}e^{\tilde{w}_k}}.
\end{align}
The second output of the layer is a set of trainable signals $\{\hat{\beta}_k(t)\}_{k=1}^{K}$ that represents some noise realizations in time domain. For each of these realizations, we construct the Hamiltonian $H^{(1)}_S(t)$ using a custom whitebox implementing Equation \ref{equ:H1}. Next, the output passes through a \textit{modified} quantum evolution whitebox that calculates the modified interaction unitary $\tilde{U}_I^{(k)}(T)$ for each noise realization $\hat{\beta}_k(t)$ using Equation \ref{equ:UI_tilde}. The time-ordered evolution is similarly approximated as in Equation \ref{equ:timeorder}. The two paths related to the noise then merge into the ``$V_O$ Construction'' layer, which is a whitebox that calculates the estimate of the operator as
\begin{align}
    \hat{V}_O(T) = \frac{1}{K} O^{-1} \sum_{k=1}^{K}{w_k \tilde{U}_I^{(k)}(T)^{\dagger} O \tilde{U}_I^{(k)}(T)}.
\end{align}
This equation represents an approximation of the classical expectation in Equation \ref{equ:Vo} that is defined over a continuous distribution of all possible noise realizations, by a weighted average over a discrete distribution of $K$ realizations. These special weights and realizations are the only trainable parameters in the model. Therefore, in order to minimize the loss function (which represents the error between the predicted outputs and the actual desired outputs), the training algorithm will be forced to find the optimal values for these parameters such that distance between the actual $V_O(T)$ operator and the estimated one $\hat{V}_O(T)$ is minimized. 

The final layer in the model is the output layer which is a whitebox that calculates the quantum measurements using Equation \ref{equ:O}. The input to the layer is the estimated $\hat{V}_O(T)$ operator from the noise path, and the $U_0(T)$ from the control path. The initial state of the qubit is parameter of this layer, and so the layer can generate measurements for multiple initial states. As discussed earlier, we can use the full 18 combinations of initial states and observables. In this case, we will need separate $V_O$ construction layer for each observable connected to a quantum measurements layer. However, we do not need multiple noise or control paths, because the control unitary $U_0(T)$ and the modified interaction unitary $\tilde{U}_I(t)$ do not depend on the observables or the initial states. This is a consequence of the linearity of quantum mechanics. 

We can see the difference between the proposed structure and the original one presented in \cite{BQNS}. The original design used standard blackboxes to generate a parameterization of the $V_O$ operator. The input to the blackboxes was the control pulses. So, in some sense the model tries to learn the modified interaction picture, which encodes the interaction of control and noise. Whereas in this paper, we directly implement the modified interaction picture with suitable whiteboxes, and use a customized blackbox to generate noise parameters. Moreover, in this paper we only use the time-domain representation of the control as the model input, instead of having two inputs (the parameterization and time-domain representation) in \cite{BQNS}. This simplifies the implementation of the model. The original design in \cite{BQNS} can still be used in the protocol. However, we choose to showcase a different graybox in this paper to emphasize the idea of building physics-aware ML models. The combination of blackboxes and whiteboxes is not unique and there is flexibility in the way we partition our model. This is similar to standard ML practice. There are some basic building block structures that can be combined in different ways and with various hyperparameters to tune.

The dataset construction and the training and testing processes for the proposed graybox follows very similarly the method of \cite{BQNS}. The MSE is used as a loss function, and an experimental dataset is constructed by applying random control pulses and measuring the observables. The difference between this paper and \cite{BQNS}, is that for the noise detection problem under consideration, we need to repeat the whole procedure (dataset construction, model training, and model testing) for each noise profile. Therefore, the output of this stage is a set of $N$ trained ML grayboxes that model the spectator qubit corresponding to each possible noise profile. We do not use the same model for multiple profiles, each profile is associated with a different model. They all have the same structure, but they end up with different trained parameters because they were trained on different datasets. A final note is that this process of constructing datasets and training ML models is very lengthy. However, it is only performed once at the beginning of the protocol and not repeated during the execution phase. 
%%%%%%%%%%%%%%%%%%%%%%%%%%%
\subsubsection{Stage II: Feature Extraction}\label{sec:stage2}
The outputs of the first stage of the protocol is a set of trained ML models corresponding to each possible noise profile. These models could be used to predict the measurement outcomes given the control pulses. But, they also can be used to predict the $V_O$ operators given the control pulses by simply probing the output of the $V_O$ construction layer. Moreover, the NN representation of $\{ V_{\sigma_\alpha}\}$ for each noise profile is equivalent to knowledge the $V^{(i)}.$

Therefore, we can use this ``reduced'' model as a part of an optimization routine to do quantum control tasks. As discussed earlier, the available control is crucial for the success of the detection. So, the target of the second stage of the protocol is to find the optimal discriminating control pulses that maximally separates the noise profiles. In order to do so, we need a criterion for the optimization. Here, we propose a heuristic based on the  average distance between the $V_O$ operators for each noise profile. 

To achieve this, we have to optimize over a choice of control sequence, observable and initial state. At first sight, this can be achieved by finding the pulse $\hat{\mathbf{f}}(t)$ and observable $O$ which maximizes the cost function
\begin{align}
%    \hat{\mathbf{f}}(t); O = \argmax_{\mathbf{f}(t)}
    C= \sum_{i,j \in \{1,2,\cdots N\}}\|V_O^{(i)} - V_O^{(j)}\|,
\end{align}
where $V_O^{(i)}$ is the $V_O$ operator of the $i^\text{th}$ noise profile and $\|\cdot\|$ is any matrix norm (in this paper we choose the Frobenius norm). However, experimentally only expectation values of an operator relative to an initial state, and not the $V_O$ are accessible. 

At this point there are at least two options. First, one can cycle over various preparations in order to reconstruct the $V_O$ for the optimal choices. This is the method we will numerically demonstrate in this paper. A second option, however, is to include in the optimization the state of the spectator. Notice for a fixed state, and observable one only accesses a portion of the $V_O$ operator. Indeed, writing $V_O = \sum v_O^\alpha \sigma_\alpha$ and $\rho_S(t) = r_\alpha \sigma_\alpha$ one has that 
$$
\langle O \rangle =  \sum_{\alpha,\alpha'}v_o^\alpha r_{\alpha'}  Tr[\sigma_\alpha \sigma_{\alpha'} O], 
$$
i.e., the expectation value is a specific linear combination of the projections $v_o^\alpha$ of $V_O$ on the system Hilbert space. In this scenario one then seeks the optimal state, observable and control sequence that optimize the cost function
\begin{align}
%    \hat{\mathbf{f}}(t); O = \argmax_{\mathbf{f}(t)}
    C'= \sum_{i,j \in \{1,2,\cdots N\}}\|\langle O^{(i)} \rangle - \langle O^{(j)} \rangle \|.
\end{align}
Because the second scenario is more restricted, one expects that the ability to discriminate between noise profiles will be in general reduced. The benefit, however, is that it involves a single preparation procedure, which immediately implies a 4x improvement in the number of necessary experiments but could also represent an experimentally friendlier setup.  Other types of considerations can enter the optimization, for example, when one considers a fixed number of shots and thus limited statistics, and all of them will represent a trade-off between speed and quality of discrimination. Ultimately what enters the optimization should be tailored to the specific experimental scenario under consideration.

\subsubsection{Stage III: Classifier Training}\label{sec:stage3}
After the second stage of the proposed protocol is executed, we obtain the optimal control pulse that best discriminates between the different noise profiles. The next step would be to train a classifier for the noise detection. The inputs to the NN will be the measurements that correspond to the optimal control pulses. These measurements will be estimated from the trained ML models from stage 1. This is the second use of the trained models, besides using them for pulse design. For each noise profile, we use the corresponding trained model to estimate the value of the measurement. The desired output of the classifier will be the class label using one-hot encoding, i.e. the label would be a vector of all zeros except at the position of the correct profile, in which case it takes the value 1. This means the first profile will be labeled as $[1,0,\cdots 0]$, the second profile as $[0,1,0,\cdots 0]$, etc. Therefore, a training example can be defined as a pair of vectors, the first being the estimated measurements and second is the encoding of the noise profile. This implies the training set will consist of exactly $N$ examples. In practice, this is not sufficient to train a standard ML classifier. Therefore, we construct the training set of the classifier differently. For each of the $N$ basic examples, we generate $R$ replicas each with white Gaussian noise added to the measurements. In signal processing, adding artificial noise to a signal to achieve a useful target is referred to as ``dithering''. There are two reasons for performing this step. The first is as we mentioned to increase the size of the training set from $N$ to $NR$. The second is to model actual errors that we would encounter in the testing phase. These errors manifest themselves as discrepancies between the predicted measurements using the trained models, and the measurements we obtain experimentally. There are two sources of such errors: 
\begin{itemize}
    \item Experimental errors such as State Preparation and Measurement (SPAM) and finite-sampling.
    \item Prediction errors due to the use of ML models.
\end{itemize}
And so to increase the robustness of the classifier, we must have noisy examples from each class, and hence the dithering step. Regarding the probability distribution of the dithering noise, there might be better distributions that takes into account models of those errors, however, this out of scope of this paper. The strength of the dithering noise has to be chosen carefully, to avoid the situation where all the classes overlap completely. At the end, the errors we are discussing should be minimal in practical situations. Note, that the dithering noise we are discussing in this context is artificial due to imperfections of the experiments and models. It has nothing to do with fundamental quantum noise that affects the evolution of the qubit, and which we aim to detect. 

For the architecture of the classifier, we choose a standard NN blackbox to build the classifier. It consists of three layers: the first has $N$ neurons, the second has $3N$ neurons, and the last layer has $N$ neurons. The two hidden layers have a hyperbolic tangent activation while the output layer has softmax activation. We use an ADAM \cite{Adam} optimization algorithm and the MSE as loss function. In general, there is a great flexibility to choose the hyperparameters defining the architecture, apart from the output layer which has to be chosen that way to generate a one-hot encoding. Once the training set is constructed for the classifier, the training is performed, and we also generate a very similar testing dataset to check the performance of the classifier.

This is the final stage of the training phase of our proposed protocol. The outcomes of this phase is the optimal control pulse sequence that best discriminated between the different noise profiles, and a trained classifier that acts on the corresponding optimal measurements. All these stages are performed at the beginning before the actual operation of the device. At this point we are ready to move on the testing phase. 
%%%%%%%%%%%%%%%%%%%%%%%%%%%%%%
\subsubsection{Stage IV: Testing}\label{sec:stage4}
The testing phase of the protocol consists of one stage that is repeated periodically. The steps involved are very efficient and thus can be executed in real-time as opposed to the training phase that requires extensive characterization as well as computations. The first step is to experimentally measure the spectator qubit using the control pulses and initial states obtained from the training phase. This will constitute a feature vector that is then passed to the trained classifier. Although, the classifier is originally trained on simulated measurements, because of the dithering step, we still expect it to perform adequately. The output of the classifier would be a probability distribution of the different noise profiles given the measurements. In the ideal case, we would expect the distribution to be completely concentrated at the correct label. In practice, this might not happen, and we may end up with a broader distribution that still peaks around the correct class. This is still perfectly fine, since we can simply infer the class label as 
\begin{align}
    \hat{n}_D =  \argmax_{i \in \{1,2,\cdots N\}} \hat{y}_i,
\end{align}
where $[\hat{y}_1, \hat{y}_2, \cdots \hat{y}_N]$ is the output of the classifier. The undesired situation is that when two or more entries have exactly the same value. In this case, the prediction would be chosen randomly between the corresponding noise profiles. This will result probably in a misclassification, which is in fact naturally expected for any detection system. In terms of quantum noise, this means that we cannot differentiate between two or more noise profiles. This implies one of the following possibilities:
\begin{itemize}
    \item The noise profiles are actually extremely close to each other, that we could consider them in fact one profile without a noticeable effect on the performance of the device.
    \item The noise profiles might still be different, but under the control constraints and capabilities available in the experiment, there is no way to practically differentiate between them.
\end{itemize}
In either case, it is impossible to distinguish between the noise profiles unless we change the control. By control constraints we mean maximum bandwidth, maximum amplitude, minimum pulse width, etc. These are imposed by the available hardware in the experiment, and they directly affect every stage in the protocol. When we perform the characterization, we use pulses having these constraints. Also, when we optimize to find the discriminating pulses, we must impose the constraints so that it is possible to implement them experimentally in the testing stage. Therefore, we have to consider the separability of noise profiles in the context of available control. We will give an example of a numerical simulation of this situation later in this paper. 

Once, the noise profile is detected, we can use the fixed map $\mathcal{F}$ as discussed earlier to infer the noise profile that is affecting the data qubit. We can then instantaneously load the proper pulse sequence that implements a desired gate of the data qubit using the pre-built lookup table $\mathcal{L}$. This ends the stage, which can then be repeated periodically. The data qubit is never interrupted while executing the gates, the spectator is used instead, fulfilling the objective of the proposed protocol. The performance of the classifier will be determined completely by the spectator qubit, under the assumption that the map $\mathcal{F}$ is fixed. If the spectator qubit is of low-quality, then it might be an advantage because in this case it might be more sensitive to the quantum noise in the environment. In the next section, we will discuss the numerical simulations that supports the presented ideas. 
%%%%%%%%%%%%%%%%%%%%%%%%%%%%%%%%%%%%%%%%%%%%%%%%%%%%%%%%%%%%%%%%%%%%%%%%%%%
\section{Simulation Results}\label{sec:sim}
In this section, we show details about the numerical simulations that were performed to demonstrate the proposed protocol. We implemented the numerical experiments in this paper using Python and Tensorflow \cite{tensorflow} and Keras \cite{keras}. The source code is publicly available as well as the datasets and the trained models that were used to generate the results in this paper \footnote{\url{https://github.com/akramyoussry/QFEND}}.
We will focus on the simulations of the stages of the protocol related to the spectator qubit only and not the data qubit. The section starts with an overview on the implementation details including the different simulation parameters and particularities of the protocol for the training phase. Next, we present the results of the testing phase by which we can assess the performance of the overall protocol. We end the section with a discussion on the significance of the obtained numerical results. 

\subsection{Training Phase}
\subsubsection{Stage I}
The first stage in the proposed protocol aims to construct a set of characterization data for each possible noise profile, and train a corresponding ML graybox. The dataset will consist of pairs of random control pulses and corresponding measurement outcomes. We will simulate this process numerically. So, we choose a Hamiltonian for the spectator qubit that is a single-axis dephasing in the form
\begin{align}
    H(t) = \frac{1}{2} f_x(t) \sigma_x + \frac{1}{2} \left(\Omega + \beta(t) \right)\sigma_z,
\end{align}
where $f_x(t)$ is the control pulses, $\Omega$ is the qubit energy gap, and $\beta(t)$ is the noise process. The evolution time interval is fixed to be $[0,T]$, and is discretized into $M$ steps. The values of the simulation parameters are $T=1$ , $M=1024$, and $\Omega=12$.  For the quantum observables, we choose the Pauli $X$, and with three positive eigenstates of the Pauli operators as an initial states. Thus, we have only three measurements to perform. We use the same Monte Carlo based technique for the simulating the open system dynamics as detailed in \cite{BQNS}. The number of noise realizations over which we take average is $K=2000$, chosen to ensure accuracy of the simulation and that enough statistics are gathered to accurately estimate a given observable's expectation value. 

For the noise profiles, we are going to generate realizations for six different random processes and we will choose different subsets to showcase the different possibilities that could occur in an  experiment. In a given experiment, the noise profiles can have common characteristics which may simplify the classification process, e.g., if all spectra are of the form  $1/f^\alpha$ plus a high frequency peak with variable position. Here we choose various types of noise profiles to showcase the flexibility of our approach. 

The six profiles are as follows:
\begin{enumerate}
    \item N0: noiseless (i.e. $\beta(t)=0$)
    \item N1: $\beta(t)$ is defined via its PSD, which take the form of a $1/f$ noise with a cutoff followed by a Gaussian bump. This can be expressed as $S_Z(f) = \frac{1}{f+1} u(15-f) + \frac{1}{16}u(f-15) + 0.5e^{-(f-30)^2/50}$, where $u(\cdot)$ is the unit step function.
    \item N2: $\beta(t)$ is a stationary Gaussian colored noise defined via its autocorrelation matrix. The coloring of the noise is simulated by performing a convolution of a white Gaussian noise signal with some deterministic signal. In particular,
    \begin{align*}
        \beta(t) = \frac{1}{10} \eta(t) \ast h(t):=\int_{-\infty}^{\infty} \eta(\tau) h(t-\tau) d\tau, 
    \end{align*}
    where $\eta(t)$ is a standard Gaussian random process with zero mean, and unit variance, and the deterministic function $h(t)$ is a fixed pulse of length $T/4$:
    \begin{align*}
        h(t) =  \begin{cases}
        1, & \text{for } 0\leq t \leq \frac{T}{4}\\
        0, & \text{otherwise}
        \end{cases}
    \end{align*}

    \item N3: $\beta(t)$ is a non-stationary Gaussian colored noise defined via its autocorrelation matrix. The non-stationarity is simulated by multiplying the stationary noise by some deterministic signal in time domain, that is 
    
    \begin{align*}
        \beta(t) =\frac{1}{5} g(t) \big(\eta(t)\ast h(t)\big) 
    \end{align*}
    where $\eta(t)$ and $h(t)$ are defined as in N2, while $g(t)$ is a triangular pulse centered at $T/2$,
    \begin{align*}
        g(t) = \begin{cases}
         T-2\left|t-\frac{T}{2}\right|, & \text{for } 0\leq t \leq T\\
        0, & \text{otherwise}
        \end{cases}
    \end{align*}

    \item N4: $\beta(t)$ is a non-stationary non-Gaussian colored noise defined via its autocorrelation matrix. The non-Gaussianity is simulated by applying a non-linear function to a Gaussian noise (in this paper we choose squaring), so
    
    \begin{align*}
        \beta(t) =  \bigg( \frac{1}{10} g(t) \big(\eta(t)\ast h(t)\big)\bigg)^2,
    \end{align*}
    where $\eta(t), h(t), $ and $g(t)$ are the same functions defined in the N2 and N3 profiles.

    \item N5: $\beta(t)$ is almost identical with the $N1$ profile, the only difference is the a slight shift in the location of the bump. The PSD is given by $S_Z(f) = \frac{1}{f+1} u(15-f) + \frac{1}{16}u(f-15) + 0.5e^{-(f-40)^2/50}$.
\end{enumerate}

For each of these noise profiles, we will create a dataset that consists of 10000 examples, where one example is a pair of a control pulse sequence, and the corresponding quantum measurements. The control takes the form of a train of Gaussian pulses of fixed width, and random amplitude and position in the form
\begin{align}
    f_x(t) = \sum_{k=1}^{n} { A_k e^{-\frac{(t-\mu_k)^2}{2\sigma^2}}},
\end{align}
where $n=5$, $\sigma=\frac{1}{6}\frac{T}{2n}$, $A_k$ is chosen randomly in the interval $[-100, 100]$, and $\mu_k$ is chosen randomly such that there are no overlapping pulses. The simulation parameters were chosen such that the qubit is affected by decoherence at the final evolution time $T$ at which the measurements are performed, but not so strongly that it is completely lost, i.e., a completely mixed state, given the available control.

After the datasets are created, simulating the collection of experimental characterization data, we train a separate ML model on each dataset. We select 9000 examples for training and 1000 examples for testing. The number of training iterations is 1000. Supplementary Figure \ref{fig:mse} shows the MSE evaluated over the training and testing examples as a function of the iteration number.

\subsubsection{Stage II}
The second stage after training the ML models for each noise profile is finding the optimal discriminating pulse sequence. Here we introduce three different scenarios, where we want to discriminate between $N=5$ profiles as follows:
\begin{enumerate}
    \item Scenario 1: The noise profiles are highly separable. In this case we use the N0, N1, N2, N3, and N4 profiles. 
    \item Scenario 2: Some of the noise profiles are close. Here we use the N5, N1, N2, N3, and N4 profiles.
    \item Scenario 3: The noise profiles are highly separable, but the control is limited to the range $[-1,1]$. Similar to Scenario 1, we use the N0, N1, N2, N3, and N4 profiles.
\end{enumerate}
For each of these three scenarios, we run the pulse optimizer for 500 iterations. The optimal pulse sequence we obtain for each scenario is shown in Figure \ref{fig:pulses}. The numerical experiments show in fact that these optimal pulses are not unique. If we run the optimizer multiple times, we can get different pulses.

\subsubsection{Stage III}
After we obtain the optimal control pulses, we proceed to final stage in the training phase of the protocol which is training the classifier on simulated optimal measurements. The first step is to use the trained ML models from stage 1 to simulate the three outcomes when we the input is the optimal pulses. These outcomes are the components of the feature vector. Next, we construct the dithered dataset by generating $R=10000$ noisy replicas of the feature vector for each class. Since we have five profiles to distinguish, the total number of examples in the dataset is 50000. The examples are then randomly split into training and testing with a split ratio of 0.1. The classifier is then trained for 500 iterations. 

This procedure is repeated for each of the three scenarios discussed in the simulations of Stage 2. With this, the training phase of the protocol is concluded. The outcomes are the optimal control pulses, and the trained classifier for each of the three scenarios. 

\subsection{Testing Phase}
With the training phase of the protocol fully executed, we are ready to simulate the testing phase. This is when we can actually assess the performance of the protocol. The procedure is to experimentally measure the spectator qubit applying the optimal control pulse, and then passing the measurement outcomes to the trained classifier. Then the procedure is repeated periodically over time. 

So, there are two main elements to simulate this part of the protocol. The first is simulating the optimal measurements. Here we once more use the Monte Carlo simulation that was used to create the datasets. We emphasize that we do not use the trained ML models of stage 1 in the training phase. Additionally, we reduce the number of noise realizations from $K=2000$ to $K=1000$. In a real experiment, this corresponds to a decreasing the number of shots we average over, and thus would speed-up the sensing step of the protocol. This is desired because at this stage, the protocol should operate in real-time. Moreover, this decreases the accuracy of the measurement, and thus it will act a good test to the robustness of the trained classifier to artificial noise. On the other hand, when we perform the characterization of the device at the beginning, we sacrifice the time in order to get high quality datasets to enhance the performance of the ML graybox models.  

The second element is simulating the process of the noise profile changing in time. For this, we generate a random sequence of integers $i_1, i_2, \cdots i_L$ of length $L=10000$, where $i_k$ denotes the index of the current noise profile. Then we simply loop over each index, generate the set of noise realizations corresponding to this profile, and run the quantum simulations as described previously. Different noise realizations are generated on the fly at each step in the sequence. This ensures there is enough randomness to mimic an actual experiment. After we run the simulation and get the measurement outcomes, we pass it to the trained classifier and store the predicted profile. 

\begin{figure}
    \centering
    \subfloat[Scenario 1]{\includegraphics[scale=0.6]{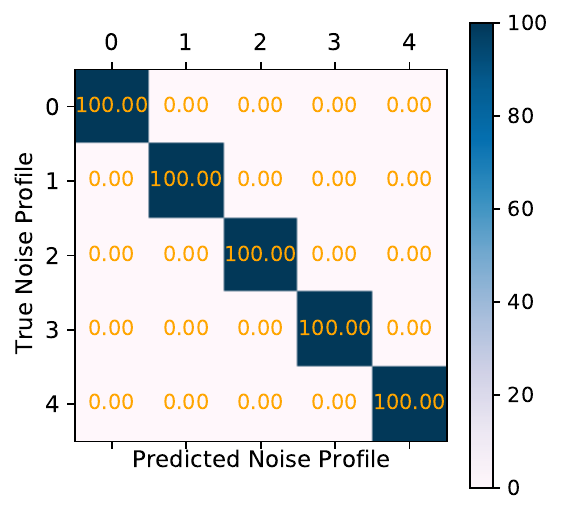}}
    \subfloat[Scenario 2]{\includegraphics[scale=0.6]{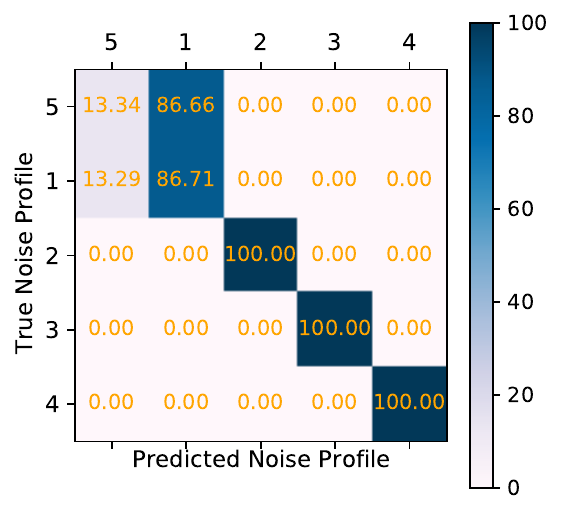}}
    \subfloat[Scenario 3]{\includegraphics[scale=0.6]{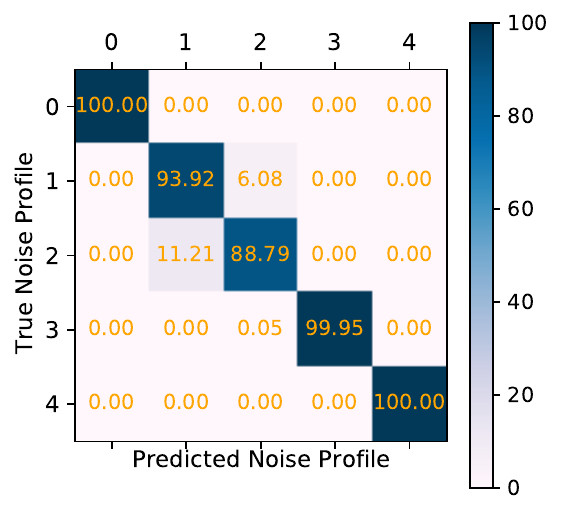}}
    \caption{The confusion matrix for the three scenarios evaluated over $10^4$ examples. The first scenario is when the noise profiles are highly separable, the second is when some profiles are close and the control is unlimited, and the final scenario is when some profiles are close and the control is limited. Each row in the confusion matrix represents the ground truth class, while the columns represent the predictions by the classifier. Each entry is the corresponding percentage of times the classifier predicted a particular class given the ground truth.}
    \label{fig:confusion}
\end{figure}

For each scenario, we do the aforementioned procedure, and then we calculate the confusion matrix as a metric for the performance of the protocol for this scenario. This is a widely used metric in ML to assess the performance of classification algorithms. The confusion matrix $C$ is an $N \times N$ matrix, where the element $C_{ij}$ is defined to be the percentage of times the classifier predicted the label $j$ whereas the ground truth label is $i$. Thus, the sum of any row should be 100\%. The best case is when the confusion matrix is a diagonal matrix with entries of 100\%. In other cases, the confusion matrix can be helpful as it can show which classes are getting mixed by the classifier. Figure \ref{fig:confusion} shows the confusion matrix for the three scenarios.

\subsection{Discussion}
The numerical simulations in this section show a promising performance for the proposed protocol. There are two main results to explore. The first is the performance of the proposed graybox ML model of the qubit. We can see from the plot of the MSE curves in Figure \ref{fig:mse} that the model is able to learn the training examples, reflected by the training MSE curve decreasing with iterations. Additionally, the model is able to generalize as demonstrated by the testing  MSE curve also decreasing with iterations, and not being significantly far from the training MSE curve. This was the case for the six noise profiles, which shows the ability of the model to learn diverse types of quantum noise.

The second results is the performance of the classifier as depicted by the confusion matrix in Figure \ref{fig:confusion}. This actually reflects the performance of the whole protocol, since the results do not depend solely on the classifier design, but rather on all other steps. In Scenario 1, the profile N0, N1, N2, N3, and N4 were used, with the control pulses allowed to have the full range. We can see that the protocol was able to successfully classify all the labels correctly, and so the confusion matrix was diagonal with 100\%. This means that the noise profiles are separated given the optimal control so there were no misclassifications. The result is also significant because the classifier is trained on the predicted measurements from the graybox, but tested on actual simulated measurements, and yet it succeeded in the task.

In Scenario 2, we have the profiles N5, N1, N2, N3, and N4. Now, profiles N5 and N1 are very close to each other. We see from the confusion matrix that the classifier was able to correctly classify all profiles, but there were misclassifications between those two profiles. This is exactly the expected behavior. This implies that under the constraints of the control pulses, these two profiles almost identical. It is also interesting that the confusion does not propagate into other classes, it is confined between N5 and N1 cases, with bias towards classifying both N5 and N1 examples to N1. Importantly, notice that the closeness between N5 and N1 given the control constraints implies that there is in principle no loss in performance in terms of optimal control on the data qubit. That is, the optimal implementation of a given gate for the N1 profile will be very close in performance to the N5. If this was not the case for a given control sequence, then it would make such sequence an ideal one to differentiate between the two profiles and should be used as part of our classifier.

Scenario 3 is similar to the first one, with the profiles chosen to be N0, N1, N2, N3, and N4. The difference is that the pulse optimizer now has an extra constraint that the amplitudes should be restricted to $[-1,1]$. In this case, we see a degradation in the performance of the classifier with misclassifications happening across various classes. This is an interesting result, because we know from scenario 1 that is it possible to distinguish between those profiles. However, due to the constraints of the control, it was not possible. This means that effectively some profiles become indistinguishable. This situation is very similar to standard quantum control. There is always a trade-off between the fidelity of a desired quantum gate, and the constraints of the allowed control (such as amplitude or bandwidth). This result shows that the effect of quantum noise is dependent on the control, and not just its statistical properties. This can be understood as well using the language of frames (see \cite{Behnam20}). 

\section{Conclusion}\label{sec:conc}
In this paper, we proposed a protocol for noise detection in a data-spectator qubit system. The spectator is used to sense the noise to prevent the interruption of the data qubit during execution of a quantum computation. All the complexity of the protocol is concentrated in the characterization phase, allowing a real-time execution during the quantum computations. The protocol is designed following a quantum feature engineering approach to allow the utilization of machine learning methods. We presented a complete framework consisting of a novel graybox model for feature generation, a quantum control method for feature extraction, and a classifier. The numerical simulations show a promising performance of the protocol and is consistent with intuition about the behavior in various scenarios.

There are limitations and extensions that can be explored in the future. The first and most challenging limitation is the assumption that we have labeled characterization data. In other words, we are able to associate the characterization to a particular noise profile. We have discussed the possibility of doing this in some practical situations, but it is still a complex procedure generally. This cannot be avoided because we used supervised learning methods to design the protocol. However, there is a rich literature about unsupervised learning in which we can classify examples into classes without the requirement of knowing the labels beforehand. This might work for noise detection applications because we are not interested in the label itself (as is the case in many applications such as object detection in images). Another limitation is the assumption of the known fixed mapping between the data and spectator noise profiles. It will be interesting to design protocol to characterize this map. Alternatively, it might be possible to design a graybox that takes into account both qubits so that we do not need to characterize the map separately.  

There are other limitations that come from the use of ML methods such as the requirement of large datasets for training, which can cause problems for large quantum systems. Therefore, a theoretical analysis of the optimal observables would help in reducing the amount of required measurements. In this paper, we chose a heuristic that allowed us to select a small set instead of an informationally-complete set. It would be interesting to explore theoretical tools such as quantum information theory (see \cite{huang2022learning}) to do this task. Additionally, we saw that in situations where the control is constrained, the distinguishability between noise profiles is affected. It will be an interesting theoretical extension to study this problem and understand how exactly the trade-offs occur, incorporating perhaps ideas from quantum information theory such as quantifying quantum state distinguishablity and statistical distance measures between noise profiles.
This would facilitate the design of the protocol, because if we know that one or more profiles are not distinguishable under our control constraints, then we do not need to have different graybox models and datasets for them. It is also possible to extend this protocol to multi-qubit systems, at least in principle. To mitigate the exponential explosion with increasing the number of qubits, we can make use of the fact that any multi-qubit quantum gate can be decomposed into a single- and two-qubit gates which reduces the size of the learning and optimization components of the problem, as one needs to learn the $V_O$ relative to single and two qubit control pulses and optimize a multi-qubit gate. The overall cost of this stage will generally not scale favorably, and so in principle one would like to enforce locality so that the potential number of data qubits is bounded. 

Regarding the numeric simulations, we made many choices regarding the design of the various machine learning tools. While our choices show a promising performance, there are many other possibilities that could lead to better results. It is also important to test the presented ideas on an actual experiment. A first simple test would be injecting noise artificially to an almost noiseless qubit and assess the performance of the protocol. Finally, we presented a new application for the quantum feature engineering approach which is noise detection, besides the original proposal in \cite{BQNS,youssry2022multi,Youssry_2020} for characterization and control of quantum systems. It would be interesting to explore further applications in other areas of quantum engineering.
%%%%%%%%%%%%%%%%%%%%%%%%%%%%%%%%%%%%%%%%%%%%%%%%%%%%%
\paragraph*{Acknowledgments}
Funding for this work was provided by the Australian Government via the AUSMURI grant AUSMURI000002. This research is also supported in part by the iHPC facility at UTS. AY is supported by an Australian Government Research Training Program Scholarship. This work was partially supported by the Australian Government through the Australian Research Council under the Centre of Excellence scheme (No: CE170100012).
%%%%%%%%%%%%%%%%%%%%%%%%%%%%%%%%%%%%%%%%%%%%%%%%%%%%%%%%%%
\bibliographystyle{apsrev4-1}
%merlin.mbs apsrev4-1.bst 2010-07-25 4.21a (PWD, AO, DPC) hacked
%Control: key (0)
%Control: author (72) initials jnrlst
%Control: editor formatted (1) identically to author
%Control: production of article title (-1) disabled
%Control: page (0) single
%Control: year (1) truncated
%Control: production of eprint (0) enabled
%
 
%%%%%%%%%%%%%%%%%%%%%%%%%%%%%%%%%%%%%%%%%%%%%%%%%%%%%%%%%
\appendix
\section{Supplementary Figures}
\begin{figure}[h]
    \centering
    \subfloat[N0]{\includegraphics[width=0.33\textwidth]{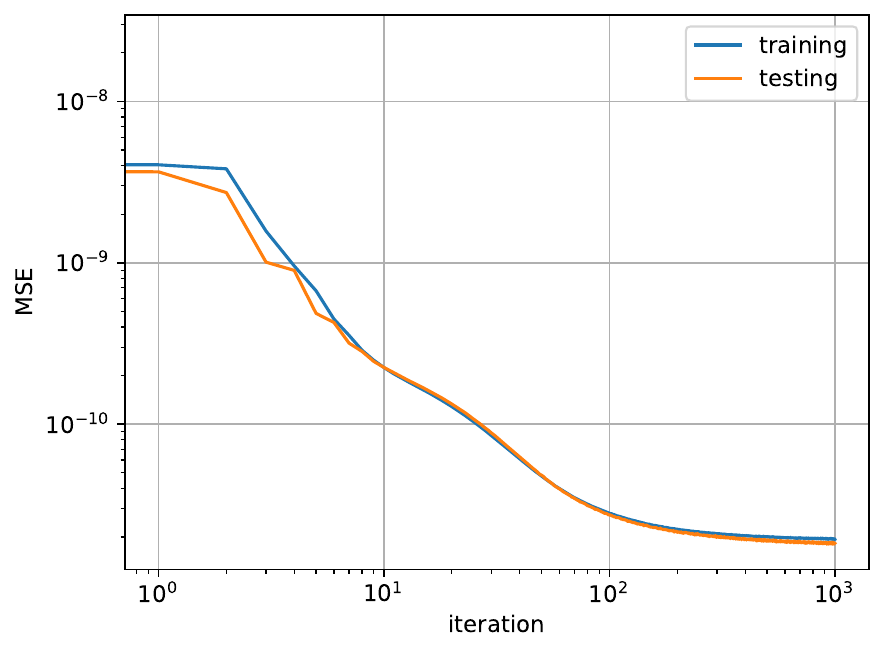}}
    \subfloat[N1]{\includegraphics[width=0.33\textwidth]{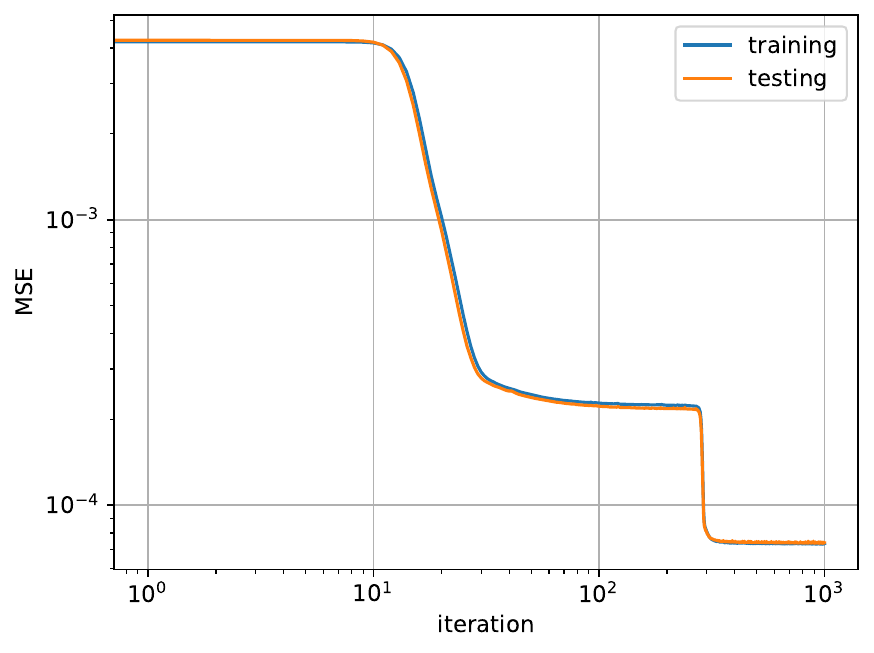}}
    \subfloat[N2]{\includegraphics[width=0.33\textwidth]{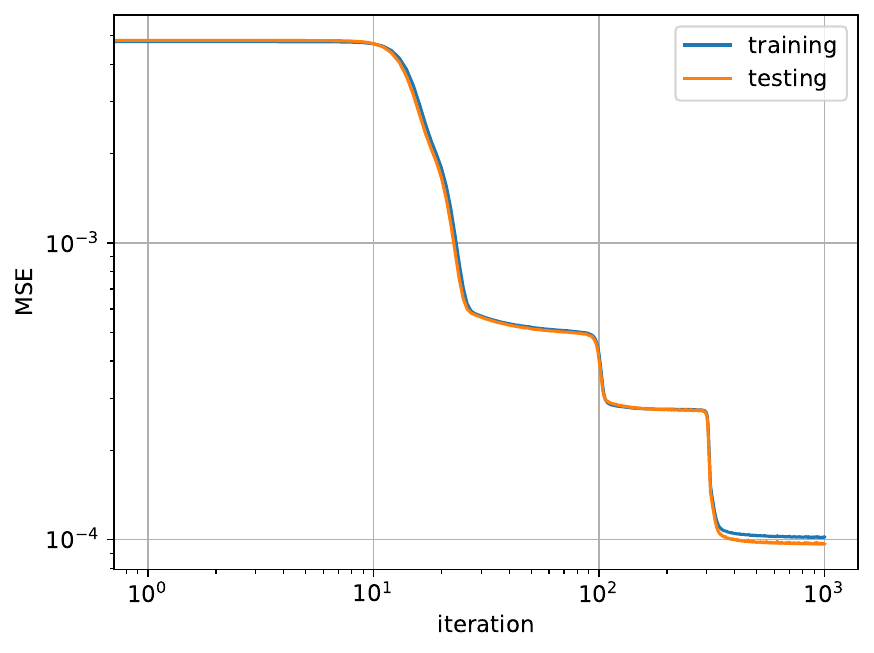}}\\
    \subfloat[N3]{\includegraphics[width=0.33\textwidth]{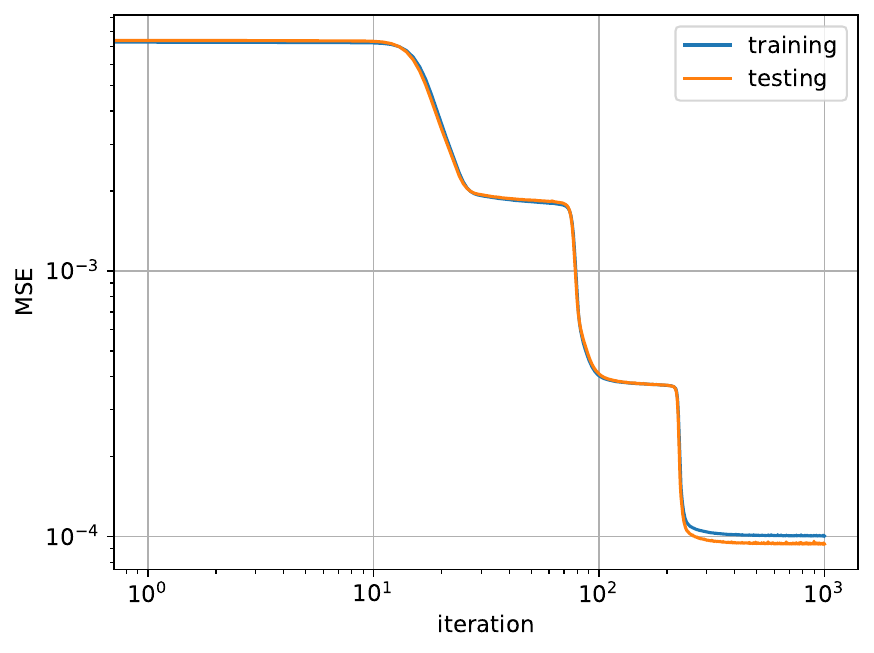}}
    \subfloat[N4]{\includegraphics[width=0.33\textwidth]{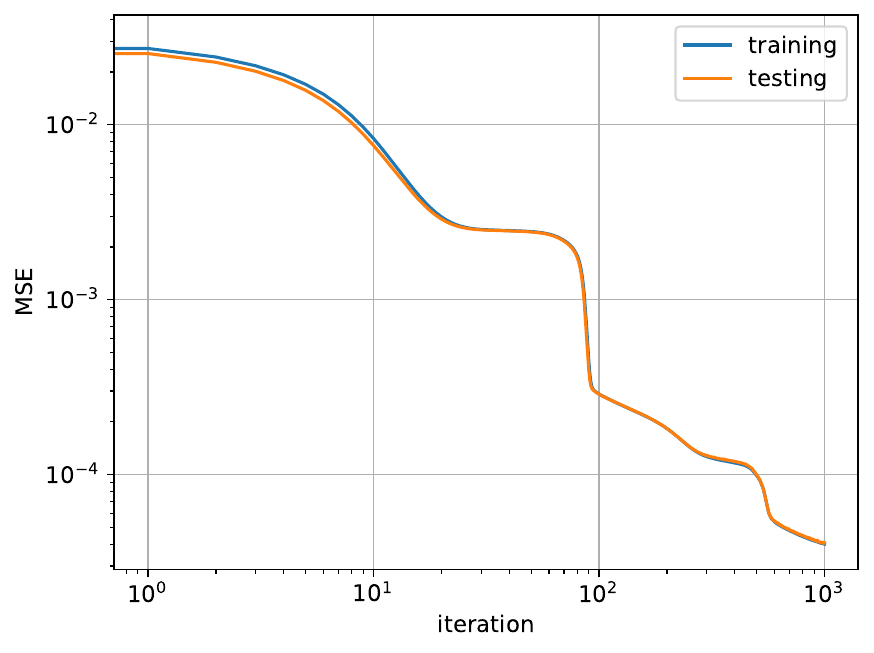}}
    \subfloat[N5]{\includegraphics[width=0.33\textwidth]{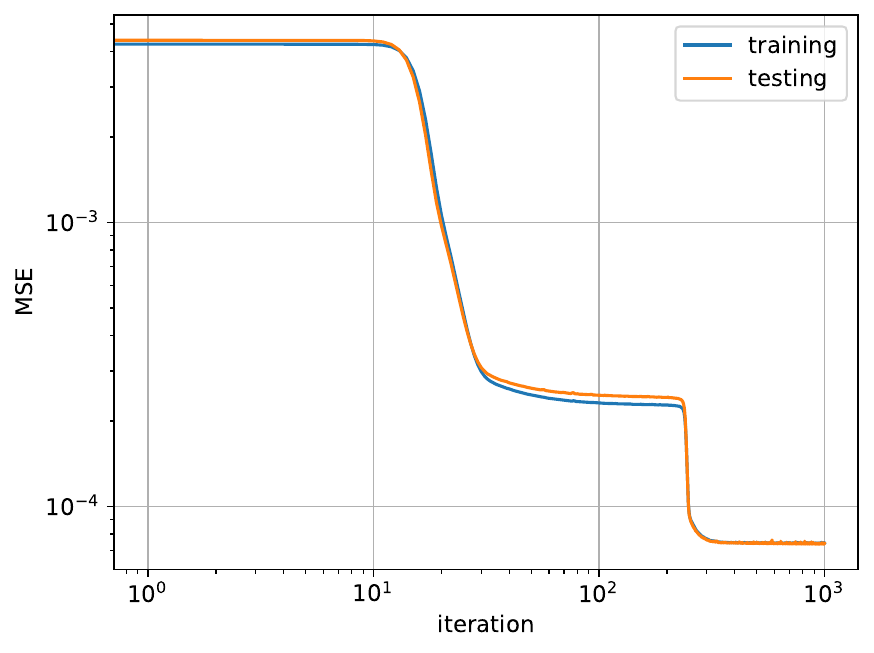}}
     \caption{The MSE evaluated for the training and testing examples for each of the simulated datasets corresponding to each noise profile.}
    \label{fig:mse}
\end{figure}

\begin{figure}
    \centering
    \subfloat[Scenario 1]{\includegraphics[width=0.33\textwidth]{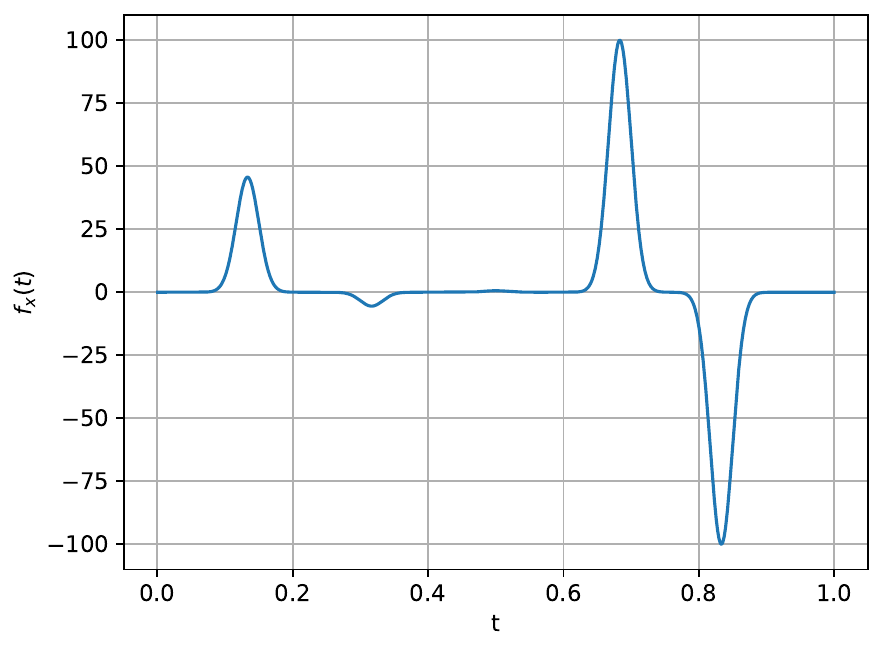}}
    \subfloat[Scenario 2]{\includegraphics[width=0.33\textwidth]{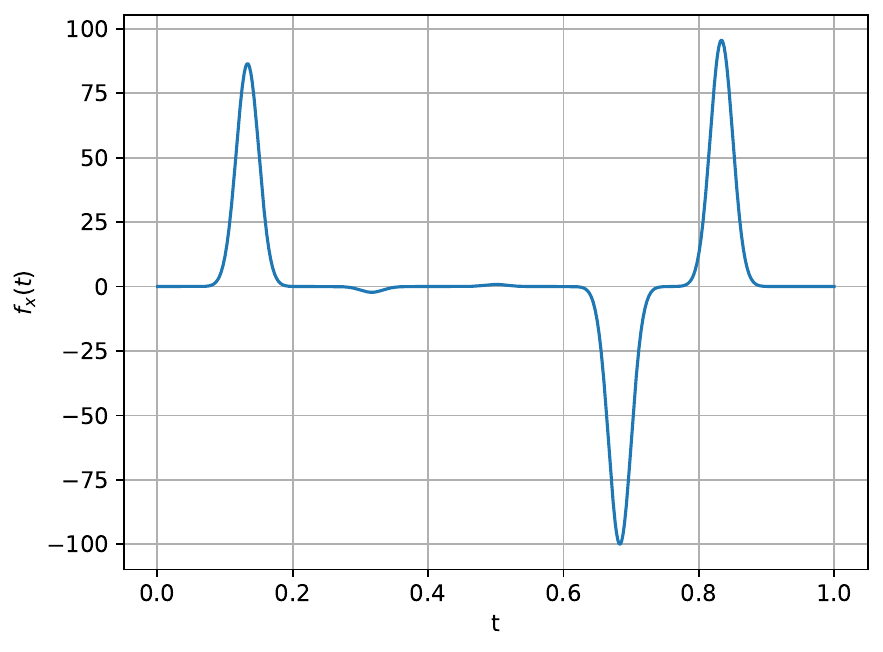}}
    \subfloat[Scenario 3]{\includegraphics[width=0.33\textwidth]{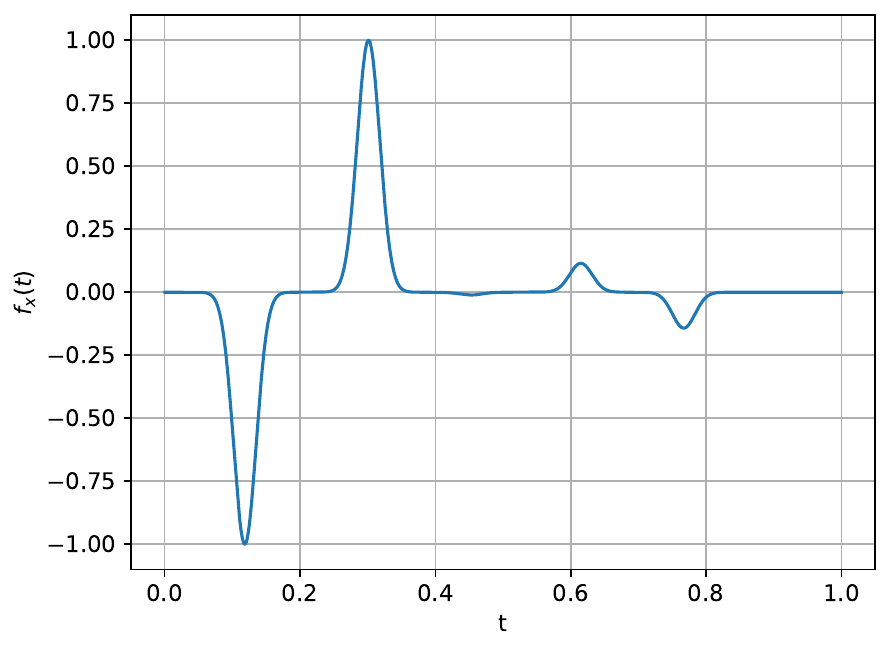}}
    \caption{The optimal discriminating pulses for the three scenarios described in the main text.}
    \label{fig:pulses}
\end{figure}

\end{document}